\documentclass[12pt,a4paper]{article}

\usepackage{a4wide}
\usepackage{amsmath}
\usepackage{bm}
\usepackage{amssymb}
\usepackage{hyperref}
\usepackage{epic}

\newcommand{\Tr}{{\rm Tr \,}}
\newcommand{\Op}{\mathcal{O}}
\newcommand{\fldZ}{\mathcal{Z}}

\newcommand{\fldD}{\mathcal{D}}

\newcommand{\alg}[1]{\mathfrak{#1}}
\newcommand{\superN}{\mathcal{N}}

\newcommand{\gym}{g\indups{YM}}
\newcommand{\indups}[1]{_{\mathrm{\scriptscriptstyle #1}}}
\newcommand{\gaba}{\gamma\inddowns{ABA}}
\newcommand{\inddowns}[1]{^{\mathrm{\scriptscriptstyle #1}}}

\newcommand{\sfrac}[2]{{\textstyle\frac{#1}{#2}}}

\newcommand{\HS}{S}
\newcommand{\M}{M}

\makeatletter
\def\mr@ignsp#1 {\ifx\:#1\@empty\else #1\expandafter\mr@ignsp\fi}%
\newcommand{\multiref}[1]{\begingroup
\xdef\mr@no@sparg{\expandafter\mr@ignsp#1 \: }%
\def\mr@comma{}%
\@for\mr@refs:=\mr@no@sparg\do{\mr@comma\def\mr@comma{,}\ref{\mr@refs}}%
\endgroup}
\makeatother

\numberwithin{equation}{section}

\begin{document}
\thispagestyle{empty}

\begin{flushright}\footnotesize
\texttt{AEI-2007-024}\\
\vspace{0.5cm}
\end{flushright}
\setcounter{footnote}{0}

\begin{center}
{\Large{\bf Dressing and Wrapping }}
\vspace{15mm}

{\sc A.~V.~Kotikov $^{a,c}$, L.~N.~Lipatov $^{b,c}$,
A.~Rej $^d$, M.~Staudacher $^d$ and V.~N.~Velizhanin $^{b,c}$}\\[5mm]

{\it $^a$ Bogoliubov Laboratory of Theoretical Physics\\
Joint Institute for Nuclear Research\\
141980 Dubna, Russia}\\[5mm]

{\it $^b$ Theoretical Physics Department\\
Petersburg Nuclear Physics Institute\\
Orlova Roscha, Gatchina\\
188300 St.~Petersburg, Russia}\\[5mm]

{\it $^c$ II.~Institut f\"ur Theoretische Physik,
     Universit\"at Hamburg\\
     Luruper Chaussee 149, D-22761 Hamburg, Germany}\\[5mm]

{\it $^d$ Max-Planck-Institut f\"ur Gravitationsphysik\\
     Albert-Einstein-Institut \\
     Am M\"uhlenberg 1, D-14476 Potsdam, Germany}\\[30mm]


\textbf{Abstract}\\[2mm]
\end{center}

\noindent{We prove that the validity of the recently proposed
dressed, {\it asymptotic} Bethe ansatz for the planar AdS/CFT
system is indeed limited at weak coupling by operator wrapping
effects. This is done by comparing the Bethe ansatz predictions for
the four-loop anomalous dimension of finite-spin
twist-two operators to BFKL constraints
from high-energy scattering amplitudes in $\superN=4$ gauge theory.
We find disagreement, which means that the ansatz breaks
down for length-two operators at four-loop order.
Our method supplies precision tools for multiple
all-loop tests of the veracity of any yet-to-be constructed
set of {\it exact} spectral equations.
Finally we present a conjecture for the {\it exact} four-loop
anomalous dimension of the family of twist-two operators,
which includes the Konishi field. 
}
\newpage

\setcounter{page}{1}
\section{Introduction and Verdict}
\label{sec:intro}

Two-dimensional integrable structures appeared for
the first time in four-dimensional gauge field theories
in the context of high-energy scattering  in QCD.
In a certain leading logarithmic approximation the
scattering amplitudes of colorless particles are well
described by the exchange of two effective particles,
termed reggeized gluons. A compound of two of these
particles is frequently called the {\it pomeron}.
In the planar limit, the associated dynamics is governed by an
integrable Hamiltonian\footnote{It was shown in
\cite{Lipatov:1993yb} that the Hamiltonian is a member of a set
of mutually commuting charges generated by a
monodromy matrix satisfying the Yang-Baxter equation.
This Padua University preprint had been submitted to
Physics Letters B and was rejected by the referee.}
\cite{Lipatov:1993yb}.
Shortly after, this Hamiltonian was identified as
the direct sum of two commuting non-compact spin-zero
Heisenberg magnets \cite{Lipatov:1994xy}. The length of this
spin chain equals the number of reggeized gluons considered.
Therefore the leading dynamics of the pomeron is
described by a very short spin chain with two sites.

Some years later integrable spin chains also resurfaced
in the analysis of planar one-loop anomalous dimensions of
composite local ``twist'' operators in QCD
\cite{Lipatov:1997vu,Braun:1998id}.
The integrable structures appearing, respectively,  in the context of
reggeization and of anomalous dimensions are frequently
confused even though the considered physical
phenomena are quite different. However, when focusing on
a more symmetric relative of QCD, the $\superN=4$
gauge theory, deep and surprising connections indeed link the
respective integrable structures \cite{Lipatov:1997vu}.
In the $\superN=4$ case, the above ``confusion'' therefore
actually expresses a profound insight.

The next step towards unravelling the exactly solvable
structure of planar $\superN=4$ gauge theory came through
the discovery that not only the sector of quasi-partonic
twist operators, but in fact the complete set of local composite
operators is described at one-loop by an integrable
$\alg{psu}(2,2|4)$ non-compact supermagnet
\cite{Minahan:2002ve}. Its spectrum
is hence described by a nested Bethe ansatz.
Much evidence was found that integrability
is {\it not} destroyed by radiative corrections, and
that the Bethe ansatz extends to higher loops
\cite{various analyses,Staudacher:2004tk,Beisert:2005tm}. This
led to a set of higher loop Bethe equations \cite{Beisert:2005fw},
which were accurate to three-loop order, but
nevertheless still incomplete at four loops and beyond,
in two distinct ways.

\begin{itemize}

\item{Firstly, the Bethe ansatz \cite{Beisert:2005fw} contained
an unknown {\it dressing} factor which was initially
introduced in order to reconcile the integrable structures
of gauge and string theory \cite{Bena:2003wd},
linked through AdS/CFT,
in certain long-operator limits \cite{Arutyunov:2004vx}.
The understanding of its necessity and structure was
subsequently refined in a series of important papers
\cite{moredressing}. Its existence was finally indirectly
proven through an impressive field-theoretic four-loop
calculation \cite{Bern:2006ew}, following a testing
procedure proposed in \cite{Eden:2006rx}.
Its precise form was written down in \cite{Beisert:2006ez}
contemporaneously with \cite{Bern:2006ew},
and agrees quantitatively with the field theory
computation to a very high precision \cite{Cachazo:2006az}.
A self-consistent
derivation of the dressing phase from first principles is still lacking.
Very recently, however, it was demonstrated that the proper
convolution structure of the phase arises from a nested Bethe ansatz
\cite{Rej:2007vm}. See also the comments in \cite{Gromov:2007cd}.
}

\item{Secondly, the Bethe ansatz \cite{Beisert:2005fw} does
not necessarily incorporate {\it wrapping} effects, as it is
by construction \cite{Staudacher:2004tk} asymptotic.
The point is that the all-loop dilatation operator of the
gauge theory is long-range, i.e.~the interactions
link at $\ell$ loop orders $\ell +1$ neighboring sites on a lattice
spanned by the partons. The definition of an S-matrix requires
an asymptotic region (see \cite{Staudacher:2004tk} for a discussion).
If the interaction range exceeds the
size of the system the asymptotic region shrinks to zero
and the Bethe ansatz might well break down. One can show
\cite{Beisert:2005fw} that in $\superN=4$ this cannot
happen up to three-loop order. A wrapping-induced breakdown
might however occur at four-loop order for the shortest
possible operators. Investigating this issue is the main
purpose of this paper.
}
\end{itemize}

The shortest possible local composite operators in the
$\superN=4$ theory
are the so-called twist-two operators. For a simple representative
of these one starts from the protected half-BPS states
$\Tr \fldZ^2$ and inserts $\M$ covariant derivatives $\fldD$:
\begin{equation}
\label{twisttwo}
\Tr \left( \fldZ\, \fldD^\M\, \fldZ\,\right) + \ldots\, .
\end{equation}
In the spin chain picture this is a non-compact $\alg{sl}(2)$
spin $=-\sfrac{1}{2}$ length-two Heisenberg magnet with $\M$ magnonic excitations.
The dots indicate the mixing of all states where the covariant
derivatives may act on any of the two fields. For each even $\M$
there is precisely one non-BPS state whose total scaling dimension is
\begin{equation}
\label{dimension}
\Delta=2+\M+\gamma(g)\, ,
\qquad {\rm with} \qquad
\gamma(g)=\sum_{\ell=1}^\infty  \gamma^{}_{2\ell}\,g^{2\ell}\, ,
\end{equation}
where $\gamma(g)$ is the anomalous part of the dimension
depending on the coupling constant
\begin{equation}
\label{convention}
g^2=\frac{\lambda}{16\,\pi^2}\, ,
\end{equation}
and $\lambda=N\, \gym^2$ is the 't Hooft coupling constant.
States with odd $\M$ do not exist in the $\alg{sl}(2)$ sector. 
The anomalous
part $\gamma(g)$ of the dimension may be reliably
computed to three-loop order $\Op(g^6)$ by the asymptotic
Bethe ansatz \cite{Staudacher:2004tk},
see \eqref{sl2eq},\eqref{dim} of chapter~\ref{sec:verdict}.
The result agrees at two-loop order with the one obtained from an
explicit field-theory calculation \cite{Kotikov:2003fb},
and at three-loop order with a solid conjecture
\cite{Kotikov:2004er} extracted by the principle of maximum
transcendentality \cite{Kotikov:2002ab} from a rigorous field
theory calculation in QCD \cite{Moch:2004pa}.
Closely related interesting properties of perturbative
anomalous dimensions, following from certain generalized relations 
of  Gribov and Lipatov, and of Drell, Levy and Yan, are discussed in 
\cite{Dokshitzer:2005bf}.

In $\superN=4$ theory the $\ell$-th loop anomalous dimension
$\gamma^{}_{2 \ell}(\M)$ is expressed through a combination of
harmonic sums of constant degree $2\,\ell+1$. These are
defined in \eqref{harmonic} below. The relationship to
the Balitsky-Fadin-Kuraev-Lipatov (BFKL) approach
\cite{Lipatov:1976zz}
for describing high energy scattering amplitudes in gauge
theory appears upon analytically continuing the
function $\gamma(g,M)$, and therefore the $\gamma^{}_{2 \ell}(M)$,
to general, complex values of $\M$. In particular, one expects
singularities at all {\it negative integer} values of  $\M$.
The first in this series of singular points corresponds
to the above mentioned pomeron at
\begin{equation}
\label{omega}
M=-1+\omega\, ,
\end{equation}
where $\omega$ should be considered small.
Notice that in BFKL physics one more commonly uses
the variable $j$ instead of $\M$. These are related through
$\M=j-2$.  Roughly speaking, in view of \eqref{twisttwo}
we could say that the BFKL
pomeron of $\superN=4$ gauge theory
is described by the non-local gauge-invariant operator
\begin{equation}
\label{pomeron}
{\rm pomeron} = \Tr\left(  \fldZ\, \fldD^{-1+\omega}\, \fldZ\,\right) .
\end{equation}
The BFKL equation relates $\gamma(g)$ and $g$ in the vicinity
of the point $\M=-1$. To leading (LO) one-loop order it reads
\begin{equation}
\label{LOBFKL}
\frac{\omega}{- 4\,g^2}=
\Psi\left(-\frac{\gamma}{2}\right)+\Psi\left(1+\frac{\gamma}{2}\right)-
2\,\Psi\left(1\right)\, ,
\end{equation}
where $\Psi(x)=\frac{d}{dx}\,\log \Gamma(x)$ is the logarithmic
derivative of Euler's Gamma function.
By expanding the $\Psi$-functions in infinite series it may
be rewritten as
\begin{equation}
\label{LOBFKL2}
\frac{\omega}{- 4\,g^2}=
\frac{2}{\gamma}
-2\sum_{k=1}^{\infty}\left(\frac{\gamma}{2}\right)^{2k}\zeta(2k+1).
\end{equation}

We are now ready to point out the crucial importance of
the BFKL equation as a testing device for any,
past or future, conjecture on the exact higher-loop spectrum of
anomalous dimensions in the $\superN=4$ model.
The point is that even though \eqref{LOBFKL} is only
the {\it one-loop} approximation to the true, currently unknown
relationship between the spin label $\M=-1+\omega$ and the
anomalous dimension $\gamma$, upon inversion of
the power series \eqref{LOBFKL2} we get an {\it all-loop}
prediction of the leading singular behavior of $\gamma$ as
a function of the deviation from the singularity at $\M=-1$ as
$\omega \rightarrow 0$.

This inversion is easily performed to arbitrary
orders of perturbation theory. To e.g.~four-loop order one finds
for the analytically continued anomalous dimension
\begin{equation}
\label{LOpoles}
\gamma=2\,\left(\frac{-4\,g^2}{\omega}\right)
-0\,\left(\frac{- 4\,g^2}{\omega}\right)^2
+0\,\left(\frac{-4\,g^2}{\omega}\right)^3
-4\,\zeta(3)\,\left(\frac{-4\,g^2}{\omega}\right)^4
\pm \ldots\, .
\end{equation}
This may now be compared to the result as obtained
from the dressed asymptotic Bethe ansatz (ABA).
In chapter \ref{sec:verdict} below we present its prediction for
the four-loop dimension at arbitrary positive integer
spin $\M$, see  table \ref{fourloop1}.
After analytic continuation to negative values of the spin $\M$,
and expanding in $\omega$ around the pole at $\M=-1$, see
\eqref{omega}, we find
\begin{equation}
\label{killerexpansion}
\gaba=2\,\left(\frac{-4\,g^2}{\omega}\right)
-0\,\left(\frac{-4\,g^2}{\omega}\right)^2
+0\,\left(\frac{-4\,g^2}{\omega}\right)^3
-2\,\frac{(-4\,g^2)^4}{\omega^7}
\pm \ldots ,
\end{equation}
where we have also restated the known results at less
than four loops. One observes {\it maximal violation} of
the BFKL prediction \eqref{LOpoles}: The leading singularity
in $\omega$ should be a pole of fourth order. Instead, we
find a seventh order pole, which is the maximum order
the analytic continuation of an harmonic sum of transcendentality
degree seven can yield. This means that
$\gaba$ cannot be the correct anomalous dimension of
finite spin $\M$ twist-two operators in $\superN=4$ gauge theory.
We have thus established that the long-range asymptotic
Bethe ansatz breaks down at four-loop order.

\section{Four-Loop Twist-Two from Bethe Ansatz}
\label{sec:verdict}

Let us now prove our claim.
The twist $J=2$  operators \eqref{twisttwo} of interest to us
sit in the $\alg{sl}(2)$ sector of the full $\alg{psu}(2,2|4)$ magnet.
The long-range asymptotic Bethe equations for twist-$J$
operators read in this sector \cite{Beisert:2005fw,Beisert:2006ez}
\begin{equation}
\label{sl2eq}
\left(\frac{x^+_k}{x^-_k}\right)^J=
\prod_{\substack{j=1\\j \neq k}}^\M
\frac{x_k^--x_j^+}{x_k^+-x_j^-}\,
\frac{1-g^2/x_k^+x_j^-}{1-g^2/x_k^-x_j^+}\,
\exp\left(2\,i\,\theta(u_k,u_j)\right),
\qquad
\prod_{k=1}^M \frac{x^+_k}{x^-_k}=1\, .
\end{equation}
These are $\M$ equations for $k=1,\ldots,\M$ Bethe roots
$u_k$, with
\begin{equation}\label{definition x}
x_k^{\pm}=x(u_k^\pm)\, ,
\qquad
u^\pm=u\pm\tfrac{i}{2}\, ,
\qquad
x(u)=\frac{u}{2}\left(1+\sqrt{1-4\,\frac{g^2}{u^2}}\right),
\end{equation}
and where the dressing phase $\theta$
is a rather intricate function conjectured in
\cite{Beisert:2006ez}. Here we will only need it to leading four-loop
order, where it reads
\begin{equation}
\label{4loopphase}
\theta(u_k,u_j) =4\,
\zeta(3)\,g^6 \big(q_2(u_k)\,q_3(u_j)-q_3(u_k)\,q_2(u_j)\big)
+\Op(g^8)\, ,
\end{equation}
and where the $q_r(u)$ are the eigenvalues of the conserved
magnon charges,  see \cite{Beisert:2005fw} for details on
this formalism. Once the $\M$ Bethe roots are determined
from \eqref{sl2eq} for the state of interest, its asymptotic
all-loop anomalous dimension is given by
\begin{equation}
\label{dim}
\gaba(g)=2\, g^2\, \sum^{\M}_{k=1}
\left(\frac{i}{x^{+}_k}-\frac{i}{x^{-}_k}\right) .
\end{equation}
The equations \eqref{sl2eq} can be solved recursively order by order
in $g$ at arbitrary values of $\M$ and $J$ once the one-loop
solution for a given state is known.

It was checked in \cite{Staudacher:2004tk} up to relatively high
values of spin, that this Bethe ansatz reproduces correctly the
two- and three-loop anomalous dimensions of the twist $J=2$
operators, which are known in terms of nested harmonic sums
as obtained in \cite{Kotikov:2003fb,Kotikov:2004er}.
However unfortunate, no analytical derivation is known at the time
of writing. It would be extremely interesting to develop tools
for solving this problem. Therefore, a priori it is even less clear
how to extract the {\it four-loop} prediction from the above
Bethe equations.

This technical problem can nevertheless be surmounted. Assuming the
maximum transcendentality principle \cite{Kotikov:2002ab} at
four-loop order one may derive the corresponding expression for the
anomalous dimension by making an appropriate ansatz with unknown
coefficients multiplying the nested harmonic sums, and subsequently
fixing these constants. The latter is done by fitting to the exact
anomalous dimensions for a sufficiently large list of specific
values of $\M$ as calculated from the Bethe ansatz.

Luckily, at one-loop the exact solution of the Baxter
equation is known \cite{Eden:2006rx} and is given by a Hahn
polynomial. Knowing the one-loop roots one can then expand equation
\eqref{sl2eq} in the coupling constant $g$ order by order in
perturbation theory. The equations for the quantum corrections to
the one-loop roots are of course linear, and thus numerically
solvable with high precision.

\begin{table}[hp!]
\begin{eqnarray}
& &\bm{4\, \HS_{-7}+6\, \HS_{7}}+ 2\,( \HS_{-3,1,3} + \HS_{-3,2,2} +
      \HS_{-3,3,1} + \HS_{-2,4,1} )
  + 3\,( -\HS_{-2,5}\nonumber\\
  &+& \HS_{-2,3,-2} ) +4\,( \HS_{-2,1,4}
- \HS_{-2,-2,-2,1} -
      \HS_{-2,1,2,-2} - \HS_{-2,2,1,-2} -
      \HS_{1,-2,1,3} \nonumber \\
&-& \HS_{1,-2,2,2} -
      \HS_{1,-2,3,1} )
  + 5\,( -\HS_{-3,4}
+ \HS_{-2,-2,-3} )
  + 6\,(- \HS_{5,-2} \nonumber\\
&+& \HS_{1,-2,4} -
      \HS_{-2,-2,1,-2} - \HS_{1,-2,-2,-2} )
  + 7\,( -\HS_{-2,-5}
+ \HS_{-3,-2,-2} \nonumber\\
&+&
      \HS_{-2,-3,-2} + \HS_{-2,-2,3} )
  + 8\,( \HS_{-4,1,2} + \HS_{-4,2,1} - \HS_{-5,-2} - \HS_{-4,3}\nonumber \\  &-&
      \HS_{-2,1,-2,-2}
+ \HS_{1,-2,1,1,-2} )
  + 9\,\HS_{3,-2,-2}
  -10\,\HS_{1,-2,2,-2}
  + 11\,\HS_{-3,2,-2}\nonumber\\
 &+& 12\,( -\HS_{-6,1} + \HS_{-2,2,-3}
+ \HS_{1,4,-2} + \!\HS_{4,-2,1} + \!
      \HS_{4,1,-2} - \!\HS_{-3,1,1,-2} -\!
      \HS_{-2,2,-2,1}\nonumber\\
&-& \!\HS_{1,1,2,3} - \!
      \HS_{1,1,3,-2} - \!\HS_{1,1,3,2} - \!
      \HS_{1,2,1,3}
- \HS_{1,2,2,-2} -
      \HS_{1,2,2,2} - \HS_{1,2,3,1} -
      \HS_{1,3,1,-2} \nonumber \\
&-& \HS_{1,3,1,2} -
      \HS_{1,3,2,1} - \HS_{2,-2,1,2} -
      \HS_{2,-2,2,1} - \HS_{2,1,1,3}
-
      \HS_{2,1,2,-2} - \HS_{2,1,2,2}\nonumber\\
&-&
      \HS_{2,1,3,1} - \HS_{2,2,1,-2} -
      \HS_{2,2,1,2} - \HS_{2,2,2,1} -
      \HS_{2,3,1,1} - \HS_{3,1,1,-2} -
      \HS_{3,1,1,2}
- \HS_{3,1,2,1}\nonumber\\
&-& \HS_{3,2,1,1} )
+ 13\,\HS_{2,-2,3}
  -14\,\HS_{2,-2,1,-2}
  + 15\,( \HS_{2,3,-2} + \HS_{3,2,-2} )
 \nonumber \\
&+& 16\,( \HS_{-4,1,-2}
+ \HS_{-2,1,-4} - \!
      \HS_{-2,-2,1,2} - \!\HS_{-2,-2,2,1} - \!
      \HS_{-2,1,-2,2} - \!\HS_{-2,1,1,-3} \nonumber\\
&-& \!
      \HS_{1,-3,1,2} - \!\HS_{1,-3,2,1} - \!
      \HS_{1,-2,-2,2}
- \HS_{2,-2,-2,1} +
      \HS_{-2,1,1,-2,1} + \HS_{1,1,-2,1,-2}\nonumber \\
 &+&
      \HS_{1,1,-2,1,2} + \HS_{1,1,-2,2,1} )
  -17\,\HS_{-5,2}
  + 18\,( -\HS_{4,-3}
- \HS_{6,1} + \HS_{1,-3,3} )\nonumber\\
  &+& 20\,( -\HS_{1,-6} - \HS_{1,6} -
      \HS_{4,3} + \HS_{-5,1,1} +
      \HS_{-4,-2,1} + \HS_{-3,-2,2} +
      \HS_{-2,-4,1} \nonumber\\
&+& \HS_{-2,-3,2} +
      \HS_{1,3,3} + \HS_{3,1,3} +
      \HS_{3,3,1} - \HS_{1,1,-2,3} -
      \HS_{1,2,-2,-2} - \HS_{2,1,-2,-2} )\nonumber \\
 &-&21\,\HS_{3,4}
+ 22\,( \HS_{1,-2,-4} + \HS_{2,2,3} +
      \HS_{2,3,2} + \HS_{3,-2,2} + \HS_{3,2,2})
+ 23\,( -\HS_{-3,-4} \nonumber \\
&-& \HS_{5,2} +
      \HS_{2,-2,-3} )
+ 24\,( -\HS_{-4,-3} + \HS_{1,-4,-2} -
      \HS_{1,-3,1,-2} - \HS_{1,1,1,4} -
      \HS_{1,1,4,1}\nonumber\\
&-& \HS_{1,3,-2,1} -
      \HS_{1,4,1,1} - \HS_{3,-2,1,1}
- \HS_{3,1,-2,1} - \HS_{4,1,1,1} +
      \HS_{-2,-2,1,1,1} + \HS_{-2,1,-2,1,1}\nonumber\\
&+&
      \HS_{1,-2,-2,1,1} + \HS_{1,-2,1,-2,1} +
      \HS_{1,1,-2,-2,1}
+ \HS_{1,1,1,-2,-2} +
      \HS_{1,1,2,-2,1} + \HS_{1,2,1,-2,1}\nonumber\\
&+&
      \HS_{2,1,1,-2,1} )
+ 25\,\HS_{2,-3,-2}
  + 26\,( -\HS_{2,5} + \HS_{1,4,2}
+ \HS_{2,4,1} + \HS_{4,1,2} + \HS_{4,2,1})\nonumber\\
&+& 28\,( \HS_{1,2,4} + \HS_{2,1,4} -
      \HS_{-3,1,-2,1} - \HS_{-2,1,-3,1} -
      \HS_{1,-2,1,-3} )
+ 30\,\HS_{-3,1,-3} \nonumber\\
  &+& 32\,( \HS_{1,5,1} + \HS_{5,1,1} -
      \HS_{-3,-2,1,1} - \HS_{-2,-3,1,1} -
      \HS_{1,-3,-2,1} - \HS_{1,-2,-3,1} \nonumber\\
&-&\HS_{2,2,-2,1} + \HS_{1,2,-2,1,1} +
      \HS_{2,1,-2,1,1} - \HS_{1,1,1,-2,1,1} )
  + 36\,( \HS_{1,1,5} + \HS_{1,3,-3} \nonumber\\
&+&      \HS_{3,1,-3}
- \HS_{1,1,-3,-2} - \!
      \HS_{1,1,-2,-3} - \!\HS_{1,1,2,-3} - \!
      \HS_{1,2,-2,2} - \!\HS_{1,2,1,-3} - \!
      \HS_{2,1,-2,2} \nonumber \\
&-& \!\HS_{2,1,1,-3} )
  + \!38\,\HS_{-3,-3,1}
+ 40\,( -\HS_{1,-4,1,1} - \HS_{2,-3,1,1} +
      \HS_{1,1,1,-2,2} )\nonumber\\
&-&41\,\HS_{3,-4}
  + 42\,( -\HS_{2,-5} + \HS_{1,-4,2} +
      \HS_{1,-3,-3} )
+ 44\,( \HS_{1,-5,1} + \HS_{2,-3,2}
+      \HS_{3,-3,1} )\nonumber \\
&+& 46\,\HS_{2,2,-3}
  + 48\,\HS_{1,1,-3,1,1}
  + 60\,( \HS_{1,1,-5} - \HS_{1,1,-3,2} )
+ 62\,\HS_{2,-4,1}
+ 64\,\HS_{1,1,1,-3,1}\nonumber\\
 &+& 68\,( \HS_{1,2,-4} + \HS_{2,1,-4} -
      \HS_{1,2,-3,1}- \HS_{2,1,-3,1} )
-72\,\HS_{1,1,1,-4}
-80\,\HS_{1,1,-4,1}\nonumber\\
&-&\bm{\mathbf{\zeta(3)}\HS_1(\HS_3-\HS_{-3}+2\,\HS_{-2,1})}.
\nonumber
\end{eqnarray}
\caption{The result for the four-loop asymptotic dimension
$\frac{\gaba_8 (\M)}{256}$ .
The harmonic sums are functions of $\M$ and are defined in
\eqref{harmonic}. The basis is canonical, except for the terms
stemming from the dressing factor in the last line.
}\label{fourloop1}
\end{table}

Under the further assumption that no index equalling $-1$ may appear
in the nested harmonic sums (see \cite{Kotikov:2002ab} and discussion
therein) there are, in principle, 238 terms which may potentially
contribute to the four-loop dimension. One thus needs to solve the
Bethe equations for 238 different values of spin $\M$. In order to
find the {\it exact} coefficients in front of the harmonic sums,
which fortunately are integers, it is crucial to determine at each
value of $\M$ the anomalous dimension as a numerically exact
rational number. One is thus forced to calculate with a very high
numerical precision, i.e.~one needs typically more than 1000 digits
at four loop order. It is possible to reduce the number of the terms
in the ansatz by going to a non-canonical basis of harmonic sums
\cite{Vermaseren:1998uu}. In the end one needs to determine around
170 values of $\M$ from the Bethe equations. An important trick is
to also use the information for odd values of $\M$, even though
these are unphysical. After much effort the expression given in
table \ref{fourloop1} was found. There we use the following
definition of the harmonic sums \cite{Vermaseren:1998uu}
\begin{equation}
\label{harmonic}
S_a (M)=\sum^{M}_{j=1} \frac{(\mbox{sgn}(a))^{j}}{j^a}\, , \qquad
S_{a_1,\ldots,a_n}(M)=\sum^{M}_{j=1} \frac{(\mbox{sgn}(a_1))^{j}}{j^{a_1}}
\,S_{a_2,\ldots,a_n}(j)\, .
\end{equation}
The degree of an harmonic sum is defined to be $|a_1|+ \ldots
|a_n|$. Notice that the total degree of each term in table
\ref{fourloop1} is seven in accordance with the maximal
transcendentality principle. The expressions for the finite $\M$
one-, two- and three-loop results are not reprinted here, they may
be found in \cite{Kotikov:2004er}. We have highlighted the terms in
the last line, containing the number $\zeta(3)$ induced by the
dressing factor \eqref{4loopphase}. Using~\cite{Blumlein:2003gb},
we have also rewritten the result in an interesting non-canonical 
basis, see table \ref{fourloop2} in appendix~\ref{app:asymptotic}.

We should stress that our method, which might appear to be
only approximately valid at first sight, actually leads to the {\it exact}
perturbative solution of the Bethe equations.
The reason is that a proper set of harmonic sums
spans a linearly independent basis in a finite
dimensional vector space \cite{Vermaseren:1998uu}. A wrong ansatz
produces incredibly complicated coefficients multiplying the
harmonic sums, and breaks down immediately when compared
with a further value of $\M$ which was not yet matched.

We are now ready to analytically continue the expression in
table \ref{fourloop1} to the vicinity of the pomeron pole at
$\M=-1+\omega$. 
An explanation for how this is done may be found in
\cite{Kotikov:2005gr}. It is based on a method suggested in
\cite{Kazakov:1987jk}, see also \cite{prudnikov},
\cite{Lopez:1980dj}.

Harmonic sums of degree seven
may lead to poles no higher than seventh order in $\omega$.
In fact, it is known that none of the sums in table \ref{fourloop1}
can produce such a high-order pole except for the two sums
$\HS_{7}$ and $\HS_{-7}$, which we have highlighted at
the beginning of the table. Their residues at $1/\omega^7$ are
of opposite sign. Thus, one immediately sees that the sum of the
two residues does {\it not} cancel. The precise statement
was already quoted in \eqref{killerexpansion}.
This proves our claim.

\section{NLO BFKL and Double-Logarithm Constraints}
\label{sec:constraints}

The asymptotic Bethe ansatz fails the BFKL constraint at four loops
already to leading order in the expansion around the pomeron
resonance singularity at $\M=-1$  through the failure to cancel
the erroneous seventh-order pole in $\omega$, {\it cf}
\eqref{killerexpansion}. The proper leading behavior
\eqref{LOpoles} should definitely be quantitatively reproduced
by any future proposal for exact spectral equations of AdS/CFT.

In fact, there are further known constraints
from $\superN=4$ high energy scattering amplitudes.
Here we will state what is known, in order to provide
precise tools for testing the validity of  any future proposal
for the exact spectrum. These highly non-trivial constraints
fall into two classes: Next-to-leading order (NLO) corrections
to the BFKL equation \eqref{LOBFKL}, and the so-called
double-log predictions. Let us begin by discussing the former.

\subsection{Two-Loop BFKL}
\label{sec:constraints1}

We discussed in chapter \ref{sec:intro} that the one-loop BFKL equation
\eqref{LOBFKL} leads to all-loop results for the leading singularities of the
analytically continued anomalous dimensions of twist-two operators at the
special value $\M=-1$. Likewise, the two-loop correction to the BFKL equation
leads to constraints on the next-to-leading corrections to the position of the
pomeron singularity near $\M=-1$ \cite{Fadin:1998py}. Luckily this two-loop
correction to the BFKL equation was worked out in the case of the $\superN=4$
supersymmetric gauge theory \cite{Kotikov:2000pm,Kotikov:2002ab}.

The  two-loop corrected BFKL equation, {\it cf} \eqref{LOBFKL}, for the
twist-two case can be written in the dimensional reduction scheme
as
\begin{equation}
\frac{\omega}{-4\,g^2} = \chi (\gamma )-g^2\,\delta (\gamma )\,,
\end{equation}
where
\begin{eqnarray}
\chi (\gamma ) &=&
\Psi\left(-\frac{\gamma}{2}\right)+\Psi\left(1+\frac{\gamma}{2}\right)-
2\,\Psi\left(1\right)\, ,\\[4mm]
\delta (\gamma ) &=&4\,\chi ^{\,\prime \prime } (\gamma )
+6\,\zeta(3)+2\,\zeta(2)\,\chi (\gamma )+4\,\chi (\gamma )\,\chi ^{\,\prime} (\gamma )  \nonumber \\[2mm]
& & -\frac{\pi^3}{\sin \frac{\pi \gamma}{2}}- 4\,\Phi \left(-\frac{\gamma}{2}
\right) -4\,\Phi \left(1+\frac{\gamma}{2} \right)\,.
\end{eqnarray}
%
The function $\Phi (\gamma )$ is given by
\begin{eqnarray}
\Phi (\gamma ) =~\sum_{k=0}^{\infty }\frac{(-1)^{k}} {(k+\gamma)^2 }\biggl[\Psi
\left(k+\gamma +1\right)-\Psi (1)\biggr]. \label{9}
\end{eqnarray}
This allows us, upon power series inversion, 
to compute the correction to the leading poles 
to arbitrary orders in $g$. In particular, the four-loop result
\eqref{LOpoles} is extended to
\begin{eqnarray}\label{tlbfkl}
\label{NLOpoles} \gamma=&\left(2+0\,\omega\right)
\left(\frac{-4\,g^2}{\omega}\right) -\left(0+0\,\omega
\right)\,\left(\frac{-4\,g^2}{\omega}\right)^2
&+\left(0+\,\zeta(3)\,\omega\right)\,\left(\frac{-4\,g^2}{\omega}\right)^3
\\
&& ~~~~
-\left(4\,\zeta(3)+\frac{5}{4}\,\zeta(4)\,\omega\right)\,\left(\frac{-4\,g^2}{\omega}\right)^4
\pm \ldots . \nonumber
\end{eqnarray}
%

\subsection{Double Logarithms}
\label{sec:constraints2}
The double-logarithmic asymptotics of the scattering amplitudes
was investigated in QED and QCD in the papers~\cite{GGLF} and~\cite{KirLi}
(see also \cite{BER}). It corresponds to summing the leading terms $\sim (\alpha \ln ^2s)^n$ in all orders of perturbation theory. 
In combination with a Mellin transformation, the 
double-logarithmic asymptotics allows to 
predict the singular part of anomalous
dimensions near the point $\M=-2$. According to the hypothesis
formulated in the articles~\cite{Kotikov:2000pm,Kotikov:2002ab}, one can 
calculate the anomalous dimension $\gamma$ near other 
non-physical points 
$M=j-2=-r$ ($r=2,3,...$) from the eigenvalue of the BFKL kernel
\begin{equation}
\label{FLOBFKL}
\frac{\omega}{- 4\,g^2}=
\Psi\left(-\frac{\gamma}{2}\right)+\Psi\left(1+
\frac{\gamma}{2}+|n|\right)-
2\,\Psi\left(1\right)\, 
\end{equation}
by pushing the total conformal spin $|n|$ to  negative integer values
$|n|=-r+1$, $r=2,3,\ldots\, ,$ 
rapidly enough at $\omega =M+r \to 0$:
\begin{equation}
|n|+r-1=C_1(r)\,\omega^2+{\cal{O}}(\omega^3)\,.
\end{equation}
Physically this corresponds to the double-logarithmic contributions 
$\sim (\alpha \mbox{ln}^2 s)^n\,s^{-r+2}$ in the
Regge limit $s\to \infty$. For even $r$ due to next-to-leading
corrections the argument of the second $\Psi$-function 
in (\ref{FLOBFKL}) is effectively shifted 
\cite{Kotikov:2000pm,Kotikov:2002ab}
\begin{equation}
1+\frac{\gamma}{2}+|n|\rightarrow 1+\frac{\gamma}{2}
+|n|
+\omega\, ,
\end{equation}
and we derive the following equation for $\gamma$
\begin{equation}
\label{dleven}
\gamma\,(2\,\omega+\gamma)=-16 g^2\, .
\end{equation}
The solution of this equation is
\begin{eqnarray}\label{dlevenp}
\gamma&=&-\omega+\omega\, \sqrt{1-\frac{16 g^2}{\omega^2}}
\nonumber \\
&=&
2\,\frac{(-4\, g^2)}{\omega}
-2\,\frac{(-4\, g^2)^2}{\omega^3}
+4\,\frac{(-4\, g^2)^3}{\omega^5}
-10\,\frac{(-4\, g^2)^4}{\omega^7}-\ldots\, .
\end{eqnarray}
%
%
%
Interestingly, our result in table \ref{fourloop1} {\it agrees}
at negative even integer values of the spin with the
$-10\,\frac{(-4\, g^2)^4}{\omega^7}$ term of this expansion.
This presumably means that the asymptotic expression is
quite ``close'' to the true result, even though it clashes 
with the singularity at $\M=-1$. We shall attempt
to improve it by brute force in the next chapter.

For odd values of $M$ in accordance with \cite{KirLi} one 
can obtain the  more complicated set of equations
\begin{eqnarray}
\tilde{\gamma}\,(2\,\omega+\tilde{\gamma} )&=&-16\, g^2- 16\, 
\frac{g^2}{\omega}\,\tilde{\gamma} _ {\rm a}\,,\nonumber\\
\tilde{\gamma}_ {\rm a}(2\,\omega+\tilde{\gamma}_ {\rm a})&=&
-8\,g^2+4\,g^2
\,\frac{d}{d\,\omega }\,\tilde{\gamma}_ {\rm a}\,,
\label{oddM}
\end{eqnarray} 
where $\tilde{\gamma} _{\rm a}$ is the anomalous dimension for an 
auxiliary operator in the adjoint representation of the gauge group
(in QCD it would carry octet color quantum numbers). 
The solution of these equations coincides with the Born result 
\begin{equation}
\tilde{\gamma} _ {\rm a} =-4\,\frac{g^2}{\omega} \,,\,\,\tilde{\gamma} 
=-8\,\frac{g^2}{\omega}\, ,
\end{equation}
corresponding to the fact that at $r=2k-1$ the
leading terms $\sim \frac{g^2}{\omega }(g/\omega )^{2n} $ are absent.

One can generalize the double-logarithmic equation (\ref{dleven}) 
for even $r=2,4,...$ to include
the corrections reproducing the three leading poles up to third order in
perturbation theory \cite{Kotikov:2004er}
\begin{eqnarray}
\label{dlnext}
\gamma\,(2\,\omega+\gamma)\ =\ &-&16\, g^2\left(1-\HS_1\,
\omega-(\HS_2+\zeta _2)\,\omega ^2\right)
-64\,g^4(\HS_2+\zeta _2-\HS_1^2)\nonumber\\
&-&4\,g^2\,
(\HS_2+\HS_{-2})\,\gamma^2\,,
\end{eqnarray}
where $\HS_i=\HS_i(r-1)$. 
This predicts the corresponding residues to fourth order 
\begin{equation}
\gamma =2\,\sum _{\ell=1}^{\infty}c_\ell(\omega)\,
\left(-4\,g^2\right)^\ell\,,
\end{equation}
where, {\it cf} v5 of the {\tt arXiv} version of \cite{Kotikov:2004er},
\begin{eqnarray}
c_1(\omega)&=&\frac{1}{\omega}-\HS_1-\omega(\zeta_2+\HS_2)+...\,,\nonumber\\
c_2(\omega)&=&-\frac{1}{\omega^3}+\frac{2\,\HS_1}{\omega^2}+
\frac{\zeta_2+\HS_2}{\omega}+...\,,\nonumber\\
c_3(\omega)&=&\frac{2}{\omega^5}-\frac{6\,\HS_1}{\omega^4}+
\frac{-4\,(\zeta_2+\HS_2)+4\,\HS_1^2+(\HS_2+\HS_{-2})}{\omega ^3}
+...\,,\nonumber\\
c_4(\omega)&=&-\frac{5}{\omega ^7}+\frac{20\,\HS_1}{\omega ^6}+
\frac{14\,(\zeta _2+\HS_2)-24\,\HS_1^2-4\,(\HS_2+\HS_{-2})}{\omega ^5}+...\,.
\label{dlp}
\end{eqnarray}
Note that for odd negative values of $M$ the generalized set of 
equations (\ref{oddM})  containing the next-to-leading corrections has 
more parameters. One can fix some of them from the known singularities
of the anomalous dimensions at $M+r=\omega \rightarrow 0$ ($r=1,\,3,...$)
\begin{equation}
\gamma =
-8g^2\left(\frac{1}{\omega}-\HS_1-\omega(\zeta_2+\HS_2)\right)
-32\,g^4\,\frac{\HS_2}{\omega}-128\,g^6\,
\left(\frac{2\HS_1}{\omega ^4}+
\frac{2\HS_2+S_{-2}-2\HS_1^2}{\omega ^3}\right),
\end{equation}
but we currently do not know how to predict the residues of the corresponding poles in the fourth order.

\section{Ad Hoc Improvement of Four-Loop Twist-Two}
\label{sec:experimental}

Here we attempt to experimentally improve the erroneous
four-loop result  of table \ref{fourloop1} as obtained by
the dressed asymptotic Bethe ansatz such that all BFKL and
double-logarithm constraints of section \ref{sec:constraints} are
satisfied. Obviously this has to be done in a way which
does not ruin the correct features of the expression in
table \ref{fourloop1}.
In particular, the improvement should not modify the large spin limit
nor violate the transcendentality principle. A seemingly natural way to
ensure this is to replace the explicit $\zeta(3)$
stemming from the dressing factor by an appropriate linear
combination of  $\zeta(3)$ and finite harmonic sums of degree three.
We found that there is indeed an attractive choice, namely
replacing in the last line of the expression in table \ref{fourloop1} $\zeta(3)$ by
\begin{eqnarray}
\label{guess}
&&
\zeta(3)\to
\frac{47}{24}\,\zeta(3) - \frac{1}{4}\HS_{-3} + \frac{3}{4}\,\HS_{-2}\,\HS_1 +
  \frac{3}{8}\,\HS_1\,\HS_2 + \frac{3}{8}\,\HS_3 + \frac{1}{6}\,\HS_{-2,1} -
  \frac{17}{24}\,\HS_{2,1}\, .
\end{eqnarray}
This alteration clearly preserves transcendentality,
and it is easy to check that the large spin limit is not
modified. In addition, the catastrophic behavior in
\eqref{killerexpansion} is now replaced by the correct one
in \eqref{LOpoles}. Furthermore, the constraints from \eqref{tlbfkl} and \eqref{dlp} are also satisfied\footnote{
In the case of negative odd $M$ the situation is unclear.}.
In fact, if we make a general ansatz for the 
replacement\footnote{Please note that we are {\it not}
proposing to improve the asymptotic Bethe ansatz as such by 
replacing the $\zeta(3)$ in the dressing phase \eqref{4loopphase}.
Our proposal only applies to an {\it ad hoc} repair of
the expression of table \ref{fourloop1} for twist-two operators,
and we currently do not know if or how this replacement can be
obtained from improved spectral equations.}
of $\zeta(3)$ in table \ref{fourloop1} by a linear combination
of the 7 structures on the r.h.s.~of \eqref{guess},
we find that they are fixed to the above specific values
by 8 constraints, while one of the constraints
serves as a highly non-trivial cross-check of our
procedure: 2 from the large $M$ limit
(known $\log M$ scaling, no $(\log M)^2$ terms),
4 from LO and NLO BFKL (known residues at $M=-1$,
namely $0/\omega^7, 0/\omega^5$, equation \eqref{tlbfkl}),
(the pole $1/\omega^6$ is automatically absent, because only 
the harmonic sum $\HS_1$ can combine with this pole, but
$\HS_1$ at zero equals zero)
and 2 from the four structures appearing in the 
residue of the $1/\omega^5$ pole of $c_4(\omega)$ in
\eqref{dlp} (the other two structures, as well as the
residues of $1/\omega^7,1/\omega^6$ are not
affected, and already reproduced by the asymptotic Bethe
ansatz).
We are hence tempted to conjecture with some confidence that
the expression of table \ref{fourloop1} improved with
the replacement \eqref{guess} is in fact the {\it exact}
four-loop anomalous dimension of the $\superN=4$ twist-two
operator series.

It is therefore interesting to spell out the four-loop anomalous 
dimension of the twist-two operator of lowest spin $\M=2$, i.e.~the
Konishi field, by using  the formula in table \ref{fourloop1} with
the replacement \eqref{guess}. One finds
\begin{equation}
\label{Konishi}
\gamma= 12\,g^2-48\,g^4+336\,g^6
-\left(\frac{5307}{2}+564\,\zeta(3)\right)\,g^8 + \ldots\, .
\end{equation}
These numbers should be compared to the ones at the
end of chapter 5 of \cite{Beisert:2006ez}.
Once again, \eqref{Konishi} is a result based on some
reasonable and self-consistent assumptions, and
we dare calling it a conjecture.

\section{Four-Loop Twist-Three from Bethe Ansatz}
\label{sec:twist3}

In this article we are mostly focusing on twist-two operators.
These are ideally suited for an analysis of the wrapping
problem. Furthermore, their precise relationship to the
BFKL equation is well established. Lastly, their spectrum
may be exactly found at one-loop order. This yields a firm
platform for higher orders of perturbation theory.
However, all this does of course not mean that operators
of higher twist are not interesting. Here we will report on a
novel {\it exact twist-three} one-loop solution in the $\superN=4$
model. This allows to find exact higher-loop
anomalous dimensions in terms of nested harmonic
sums, in close analogy with the twist-two case.
Unfortunately we were so far unable to test the analytic
continuation  of these expressions
with the BFKL and double-logarithm methods of
chapter \ref{sec:constraints}. We still feel that our result
should allow for some non-trivial tests in the future.

Twist-three operators are operators of the form
\begin{equation}
\Tr \left( D^{s_1} \fldZ\, \fldD^{s_2}\, \fldZ\ \fldD^{s_3}\,
\fldZ\,\right) + \ldots .
\end{equation}
with $s_1+s_2+s_3=\M$. Since the wrapping effects in $\alg{sl}(2)$ start
at ${\cal{O}}(g^{2L+4})$ the four-loop anomalous dimension of
twist-three operators should correctly follow from the asymptotic
Bethe ansatz \eqref{sl2eq} with $J=3$.
In this chapter we will proceed with the derivation of
the four-loop anomalous dimension of the ground state of twist-three
operators at even values of $\M$. It will be shown that the
anomalous dimension up to four-loop order can be again given in
terms of harmonic sums, similarly to the twist-two case. After
analytical continuation it will turn out, however, that the
anomalous dimension does not have a pole at $\M=-1$ and thus cannot
be checked with the BFKL equation. The validity of this result as
derived from the Bethe ansatz is therefore still an open question.

At one-loop the Baxter function for the twist-three operators
satisfies
\begin{equation}\label{eqq}
(u+\frac{i}{2})^3 Q (u+i)+(u-\frac{i}{2})^3 Q(u-i)=t(u) Q(u) \, ,
\end{equation}
with the transfer matrix given by
\begin{equation}
t(u)=2u^3+q_{2}u+q_{3} \, .
\end{equation}
Using the expansion
\begin{equation}
Q(u)=\prod^{\M}_{n=1}
(u-u_{i})=u^\M+c_{1}u^{\M-1}+c_{2}u^{\M-2}+c_{3}u^{\M-3}+... \,\, ,
\end{equation}
one can read off the corresponding charges
\begin{equation}
q_{2}=-(\M^2+2\M+\frac{3}{2})\, , \qquad \qquad q_{3}=c_{1} (2\M+1)
\, .
\end{equation}
For the unpaired states $c_{1}=-\sum^{\M}_{j=1} u_{j}=0$ and
$q_{3}=0$. One can then solve \eqref{eqq} exactly for even
values of $\M$
\begin{equation}
Q(u)={_4
F_3}\big(-\frac{\M}{2},\frac{\M}{2}+1,\frac{1}{2}+iu,\frac{1}{2}-iu,;1,
1, 1; 1 \big) \, .
\end{equation}
Thus the Baxter function $Q(u)$ is given by a Wilson polynomial. Wilson polynomials for twist-three operators were also found in the QCD
context, see \cite{Korchemsky:1994um}. It is straightforward to
derive the corresponding one-loop anomalous dimension
\begin{equation}\label{t31}
\frac{\gaba_{2}(\M)}{2}=4\, S_{1}(\frac{\M}{2}) \, .
\end{equation}
There exists one more solution of (\ref{eqq}) for $q_0=0$ which is, however, non-polynomial. This proves that all unpaired states for even $\M$ have energy given by (\ref{t31}).
We suspect that this is the lowest state, but we do not know any proof.

Similarly to the twist-two case one can derive from the Bethe ansatz a closed formula for the corresponding two, three and four-loop anomalous dimensions by assuming the transcendentality principle and making an appropriate ansatz. We found the following expressions, where all harmonic
sums have $\frac{\M}{2}$ as an argument
\begin{equation}\label{t32}
\frac{\gaba_{4} (\M)}{4}=-2 S_{3}-4 S_{1} S_{2} \, ,
\end{equation}
\begin{equation}\label{t33}
\frac{\gaba_{6}(\M)}{8}=2
S_{2}S_{3}+S_{5}+4S_{3,2}+4S_{4,1}-8S_{3,1,1}+S_{1}\bigg(4S^2_{2}+2S_{4}+8S_{3,1}\bigg)
\, ,
\end{equation}
\begin{eqnarray}\label{t34}
\frac{\gaba_{8}(\M)}{16}&=&\, \, \, S^3_{1}
\bigg(\frac{40}{3}S_{4}-\frac{32}{3}S_{3,1}\bigg)+S^2_{1}\bigg(20S_{5}-40S_{3,2}-56S_{4,1}+64S_{3,1,1}\bigg)
\nonumber \\
&&+S_{1}\bigg(7S_6+8S_{2,4}-24S_{3,3}-56S_{4,2}-40S_{5,1}-24S_{2,2,2}-16S_{2,3,1}\nonumber
\\&& +88S_{3,1,2}+88S_{3,2,1}+120
S_{4,1,1} -192 S_{3,1,1,1}-8\zeta(3)S_{3}\bigg) -\frac{56}{3} S_{3}
S_{4}\nonumber \\
&&-\frac{107}{6} S_{7}+3S_{2,5}+\frac{41}{3} S_{3,4}+\frac{1}{3}
S_{4,3}-17 S_{5,2}-\frac{20}{3} S_{6,1}-4 S_{2,2,3} \nonumber \\
&& -8 S_{2,3,2}-4 S_{2,4,1} +\frac{104}{3} S_{3,1,3}+52
S_{3,2,2}+\frac{88}{3} S_{3,3,1}+60 S_{4,1,2}\nonumber
\\
&&+60S_{4,2,1}+40
S_{5,1,1}+8 S_{2,3,1,1}-120 S_{3,1,1,2}-120 S_{3,1,2,1} \nonumber \\
&& -120 S_{3,2,1,1} -128 S_{4,1,1,1}+256 S_{3,1,1,1,1}.
\end{eqnarray}
It is also instructive to display $\gamma_8$ in the canonical basis
of harmonic sums
\begin{eqnarray}
\frac{\gaba_{8}(\M)}{16}&=&\, \, \, \frac{1}{2} S_{7}+7 S_{1,6}-5
S_{6,1}-5 S_{3,4}-29 S_{4,3}-32 S_{1,2,4}-32 S_{2,1,4}+32
S_{1,4,2}\nonumber
\\&& +4 S_{2,4,1}+36 S_{4,1,2}+36 S_{4,2,1}-24 S_{1,4,1,1}-24 S_{4,1,1,1}+15
S_{2,5}-21 S_{5,2}\nonumber \\
&&+24 S_{5,1,1}+24 S_{3,3,1}+24 S_{3,1,3}+24 S_{1,3,3}+44
S_{3,2,2}+40 S_{2,3,2}+20 S_{2,2,3}\nonumber \\
&&-24 S_{1,3,1,2}+16 S_{1,2,3,1}-24 S_{1,3,2,1}-24 S_{3,1,1,2}+16
S_{2,1,3,1}-24 S_{3,1,2,1}\nonumber \\
&&-24 S_{2,3,1,1}-24 S_{3,2,1,1}-24 S_{1,2,2,2}-24 S_{2,1,2,2}-24
S_{2,2,1,2} -24 S_{2,2,2,1}\nonumber \\
&&-40 S_{1,1,5}+80 S_{1,1,1,4}+32 S_{1,1,4,1}-16 S_{1,1,3,2}-64
S_{1,1,1,3,1}\nonumber\\
&&\bm{-8 \mathbf{\zeta(3)} (S_{1,3}+S_{3,1}-S_{4})}.
\end{eqnarray}
The term multiplied by $\zeta(3)$ is due to the dressing factor. Curiously, only positive indices in the harmonic sums appear. Because the argument of the harmonics sums in \eqref{t31}-\eqref{t34} is $\frac{\M}{2}$, there is no pole at $M=-1$ and thus these states are not captured by the BFKL equation. It would be very interesting to see whether one can predict their analytic structure at $M\leq-2$ from the double-log constraints. We would like to stress again the fact that equations \eqref{t31}-\eqref{t34}, even when the asymptotic character of the Bethe equations \eqref{sl2eq} is taken into account, should give correct anomalous dimensions. Below we present 
the two highest terms in the analytical continuation to
$\M=2\,(\omega-1) $
\begin{eqnarray}
\frac{\gaba_2}{2} &=& -\frac{4}{\omega} +
4\zeta(2)\omega+\ldots~,~~~~~~~
\frac{\gaba_4}{4} ~=~ -\frac{2}{\omega^3} + \frac{4}{\omega}
\zeta(2)+\ldots,
\nonumber \\
\frac{\gaba_6}{8} &=& -\frac{1}{\omega^5} + \frac{6}{\omega^3}
\zeta(2)+\ldots~,~~~~~~~
\frac{\gaba_8}{16} ~=~ -\frac{1}{2\omega^7} + \frac{5}{\omega^5} \zeta(2)\ldots\,.\nonumber \\
\label{A2.1}
\end{eqnarray}
Note that the leading singularity is the same for all points $M=2\left(\omega-(k+1)\right)$ and $k=0,1,\ldots$. The corresponding expansion of the total anomalous dimension up to four-loop order reads
\begin{eqnarray}\label{t3tad}
\gaba &=& -8 \, \frac{g^2}{\omega} \, \biggl[ 1-\zeta(2)\omega^2 + t
\Bigl(1-2\,\zeta(2)\omega^2\Bigr)+
 t^2 \Bigl(1-6\, \zeta(2)\omega^2\Bigr)\nonumber\\
&&+\,\,\,\, t^3 \Bigl(1-10\, \zeta(2)\omega^2\Bigr)+\ldots \, \biggr]+\ldots\,,
\label{A2.2}
\end{eqnarray}
where
$t=\frac{g^2}{\omega^2}$.
One can speculate on an all-loop generalization of \eqref{t3tad}.
A plausible form might be
\begin{equation}
\gaba=-8 \frac{g^2}{\omega} \left(\frac{1}{1-t}-\zeta(2)\frac{1+3 \,t^2}{(1-t)^2} \omega^2 +\ldots \, \right)+\ldots\, .
\end{equation}
It is interesting to note that the double-logarithmic behavior of
these states is different from the twist-two ones \eqref{dlevenp}.
\section{Outlook}
\label{sec:outlook}

Our result in table \ref{fourloop1} with the ensuing
\eqref{killerexpansion} proves
unequivocally that the spectral equations of
\cite{Beisert:2005fw,Beisert:2006ez}
for AdS/CFT are still incomplete as the BFKL prediction
\eqref{LOpoles} is not reproduced at four-loop order $\Op(g^8)$.

Our weak-coupling study is complementary to
indications that at strong coupling, i.e.~on the string side of the
AdS/CFT correspondence, the asymptotic Bethe ansatz
\cite{Beisert:2005fw,Beisert:2006ez}
is also incomplete when one considers finite size effect
of the string worldsheet. For one, it was argued that for a
specific spinning string solution carrying both angular momentum
w.r.t.~$AdS_5$ and $S^5$ as well as winding numbers,
and whose classical \cite{Arutyunov:2003za} and
one-loop \cite{Park:2005ji} energy is known, the ansatz
does not reproduce exponentially small terms in the size of the
system \cite{Schafer-Nameki:2006ey}. A second indication
comes from a study of the finite size effects \cite{Arutyunov:2006gs}
on the dispersion law of classical giant magnons
\cite{Hofman:2006xt}. Again, terms which fall off exponentially
with the volume are seen which cannot easily be accounted
for by the dressed asymptotic spectral equations.

We have shown that linking the integrable structures found
in the context of high energy scattering amplitudes in $\superN=4$
theory and the ones appearing in the spectral problem
leads to very strong constraints. It was pointed out
in \cite {Kotikov:2004er} and \cite{Brower:2006ea} that 
pomeron physics and anomalous
dimensions are very naturally connected through the AdS/CFT
correspondence. However, no attention had so far been payed to
the fact that integrability will presumably allow to truly
explore these connections in a quantitative and analytic
fashion. We feel that we have made a first step in this direction.

Interestingly, the breakdown we observe is completely insensitive
to the structure of the {\it dressing} factor, which also
appears at four-loop order \cite{Beisert:2006ez}, in
contradistinction to what one might have hoped for.
Recall that this dressing phase leads to the four-loop
agreement between the Bethe ansatz and the
result of a gluon scattering amplitude in $\superN=4$
theory \cite{Bern:2006ew,Cachazo:2006az,Beisert:2006ez}.
This is not contradictory. The gluon amplitude
tests the anomalous dimension in the large spin limit
$\M \rightarrow \infty$, where it was argued that the all-loop result
leads to a universal scaling function
\cite{Eden:2006rx} (see also \cite{Belitsky:2006en}),
i.e.~one which is reached by large $\M$ scaling of the lowest state
at fixed length=twist. We may therefore choose the twist
large enough to avoid leaving the asymptotic regime.
Turning this around, we might say that the wrapping terms
should be subleading in the large spin limit.
This is reassuring, in particular since by now it appears that
the scaling function matches well \cite{Kotikov:2006ts}
the known string theory results
\cite{Gubser:2002tv}. Furthermore, the analytic structure
of the dressing phase fits well a semi-classical analysis
\cite{Dorey:2007xn}.

In this article it is proven that {\it wrapping} effects are not properly
taken into account by the existing asymptotic Bethe ansatz.
We note that this was of course never claimed otherwise
by any of the current authors,
and is actually quite expected from the way this Bethe ansatz
was initially constructed \cite{Staudacher:2004tk}.
The mechanism for the breakdown of the asymptotic
approximation should be similar to
the one discussed from a field theory standpoint
in \cite{Ambjorn:2005wa}, and from a lattice model point of view in
\cite{Rej:2005qt}.

{\bf Note added:} Our results on twist-three operators in
chapter 5 were independently obtained in the  
contemporaneous paper \cite{Beccaria:2007cn}.

 \subsection*{Acknowledgments}
A.R.~and M.S.~would like to thank Lisa Freyhult, Romuald Janik
and Stefan Zieme for useful discussions.
L.N.L. thanks Yu.~Dokshitzer for helpful discussions of
the analytic properties of the anomalous dimensions of Wilson
operators with twists two and three.
L.N.L.~holds a {\it Marie Curie Chair of Excellence}
and acknowledges the support of the European Commission.
A.V.K.~is supported in part by the Alexander von Humboldt Foundation. 
L.N.L.~is supported by 06-02-72041-MSTI-a, 
V.N.V. is supported by DFG grant No. KN 365/6-1, 
and A.V.K., L.N.L. and V.N.V. are supported by 
RFBR grants 07-02-00902-a, RSGSS-5788.2006.2.
M.S.~and A.R.~thank the Jagiellonian University in
Krak\'ow for hospitality while working on parts of this
project. 
\appendix

\section{Asymptotic Four-Loop Anomalous Dimension of Twist-Two
Operators in a Non-Canonical Basis}
\label{app:asymptotic}
%
\vspace{-0.06cm}
\begin{table}[hp!]
\begin{eqnarray}
&-&\, 8\,{\HS_{-3}}\,( 4\,{\HS_{-4}} + \HS_{-2}^2 + 2\,{\HS_{-2}}\,{\HS_2} + 2\,\HS_2^2 + 3\,{\HS_4} +
     2\,\HS_{-2,2} + 8\,\HS_{-2,1,1} ) \nonumber\\
&-&   16\,\HS_{-2,1}\,( 4\,{\HS_{-4}} + \HS_{-2}^2 -
     2\,{\HS_{-2}}\,{\HS_2} - 2\,\HS_2^2 - 5\,{\HS_4} + 6\,\HS_{-2,2} +
     8\,\HS_{-2,1,1} ) \nonumber\\
&-&  \frac{8 }{3}\,{\HS_3}\,( 36\,{\HS_{-4}} + 9\,\HS_{-2}^2 + 18\,{\HS_{-2}}\,{\HS_2} + 6\,\HS_2^2 +
       9\,{\HS_4} - 56\,\HS_{-3,1} - 50\,\HS_{-2,2}\nonumber\\
&+&
       88\,\HS_{-2,1,1} )
-  16\,{\HS_2}\,( 3\,{\HS_{-5}} + {\HS_5} -
     6\,( \HS_{-4,1} + \HS_{-3,2} +
        \HS_{-2,3} - 2\,\HS_{-3,1,1}\nonumber\\
&-&
        2\,\HS_{-2,1,2} - 2\,\HS_{-2,2,1}
+
        4\,\HS_{-2,1,1,1} )  )
   -   32\,{\HS_{-2}}\,( 4\,{\HS_{-5}} + 2\,{\HS_5} - \HS_{-4,1}\nonumber\\
&-&
     \HS_{-2,3} + \HS_{4,1} - 2\,\HS_{-2,-2,1}     )
- \frac{32 }{3}\,\HS_1^3\,( 18\,{\HS_{-4}} - 3\,\HS_{-2}^2 + 3\,{\HS_4} -
       16\,\HS_{-3,1}\nonumber\\
&-& 10\,\HS_{-2,2} +
       8\,\HS_{-2,1,1} )
-   32\,\HS_1^2\,
(
{\HS_{-2}}\,
( 5\,{\HS_{-3}}
+ 3\,{\HS_3} +
     2\,\HS_{-2,1} ) \nonumber\\
&+& 3\,{\HS_2}\,
( 3\,{\HS_{-3}} + {\HS_3} - 2\,\HS_{-2,1}     )
+15\,{\HS_{-5}} + 5\,{\HS_5}
-  2\,( 10\,\HS_{-4,1} + \HS_{-3,-2} \nonumber\\
&+&
        10\,\HS_{-3,2} +\! 7\,\HS_{-2,3} - \!
        \HS_{4,1} - \!14\,\HS_{-3,1,1} + \!
        2\,\HS_{-2,-2,1} -\!10\,\HS_{-2,1,2}
- \!10\,\HS_{-2,2,1}\nonumber\\
&+& \!12\,\HS_{-2,1,1,1} )  )
-   16\,{\HS_1}\,
(
{\HS_3}\,( 12\,{\HS_{-3}}  - 12\,\HS_{-2,1}     )
+{\HS_{-2}}\,
( 8\,{\HS_{-4}} + 6\,\HS_2^2 + 9\,{\HS_4}\nonumber\\
&-&
     12\,\HS_{-3,1} - 2\,\HS_{-2,2} )
+2{\HS_2}\,( 8\,{\HS_{-4}} + 2\,\HS_{-2}^2 + 3\,{\HS_4} -
     12\,\HS_{-3,1} - 10\,\HS_{-2,2}\nonumber\\
&+&
     16\,\HS_{-2,1,1} )
+26\,{\HS_{-6}} - 3\,\HS_{-3}^2 + \HS_{-2}^3
+ 2\,\HS_2^3 + \!2\,\HS_3^2 + \!
     3\,{\HS_6}  - \! 44\,\HS_{-5,1}\nonumber\\
&-& \!
     46\,\HS_{-4,2} - \! 46\,\HS_{-3,3}
      - \!4\,\HS_{-2,1}^2
     -\! 38\,\HS_{-2,4} + \!4\,\HS_{4,2} - \!
     8\,\HS_{5,1}
+ 80\,\HS_{-4,1,1}\nonumber\\
&+&
     8\,\HS_{-3,-2,1} + 8\,\HS_{-3,1,-2} +
     84\,\HS_{-3,1,2} + 84\,\HS_{-3,2,1}
-     8\,\HS_{-2,-2,2} + 68\,\HS_{-2,1,3} \nonumber\\
&+&
     72\,\HS_{-2,2,2} + \!68\,\HS_{-2,3,1} - \!
     8\,\HS_{4,1,1} - \!144\,\HS_{-3,1,1,1} + \!
     16\,\HS_{-2,-2,1,1}
- \!120\,\HS_{-2,1,1,2}\nonumber\\
&-& \!
     120\,\HS_{-2,1,2,1}
- 120\,\HS_{-2,2,1,1} + \!
     192\,\HS_{-2,1,1,1,1} )
-  16\,( 8\,{\HS_{-7}} +\! 9\,{\HS_7} - \!16\,\HS_{-6,1}\nonumber\\
&-& \!  6\,\HS_{-5,-2} - \!16\,\HS_{-5,2} - \!
     \HS_{-4,-3}
- 17\,\HS_{-4,3} - \!
     15\,\HS_{-3,4} - \!18\,\HS_{-2,5} - \!
     5\,\HS_{4,3}\nonumber\\
&+& \!4\,\HS_{5,2} + \!
     6\,\HS_{6,1}
+ \!32\,\HS_{-5,1,1} - \!
     6\,\HS_{-4,-2,1} + \!36\,\HS_{-4,1,2}
+
     36\,\HS_{-4,2,1}\nonumber\\
&-& 4\,\HS_{-3,-3,1} -
     2\,\HS_{-3,-2,-2} - 4\,\HS_{-3,-2,2}
+
     36\,\HS_{-3,1,3} + 40\,\HS_{-3,2,2} +
     36\,\HS_{-3,3,1} \nonumber\\
&+& 2\,\HS_{-2,-4,1} -
     8\,\HS_{-2,-3,2} + 10\,\HS_{-2,-2,3} +
     34\,\HS_{-2,1,4}
+ 36\,\HS_{-2,2,3} +
     36\,\HS_{-2,3,2}\nonumber\\
&+& 32\,\HS_{-2,4,1}
-
     4\,\HS_{4,1,2} - 4\,\HS_{4,2,1} -
     72\,\HS_{-4,1,1,1} - 80\,\HS_{-3,1,1,2}
-
     80\,\HS_{-3,1,2,1}\nonumber\\
&-& 80\,\HS_{-3,2,1,1} +
     24\,\HS_{-2,-3,1,1}
+ 4\,\HS_{-2,-2,-2,1} +
     8\,\HS_{-2,-2,1,2} + 8\,\HS_{-2,-2,2,1}\nonumber\\
&-& 8\,\HS_{-2,1,1,-3} - 72\,\HS_{-2,1,1,3} -
     80\,\HS_{-2,1,2,2}
- 72\,\HS_{-2,1,3,1} -
     8\,\HS_{-2,2,-2,1}\nonumber\\
&-& 80\,\HS_{-2,2,1,2}
-
     80\,\HS_{-2,2,2,1} - 72\,\HS_{-2,3,1,1} +
     24\,\HS_{4,1,1,1}
+ 160\,\HS_{-3,1,1,1,1}\nonumber\\
&-&
     48\,\HS_{-2,-2,1,1,1} - 16\,\HS_{-2,1,-2,1,1}
+
     160\,\HS_{-2,1,1,1,2} + 160\,\HS_{-2,1,1,2,1} \nonumber\\
&+&
     160\,\HS_{-2,1,2,1,1} + 160\,\HS_{-2,2,1,1,1} -
     320\,\HS_{-2,1,1,1,1,1} )\nonumber\\
&\bm{-}&\bm{16\, \mathbf{\zeta(3)}\HS_1(\HS_3-\HS_{-3}+2\,\HS_{-2,1})}\nonumber
\end{eqnarray}
\caption{The four-loop asymptotic dimension
$\sfrac{\gaba_8(\M)}{16}$ of table \ref{fourloop1} 
in a non-canonical basis. This isolates the $\HS_1$, 
i.e~all terms divergent as $\M \rightarrow \infty$.
}\label{fourloop2}
\end{table}
%

\end{document}